\documentstyle[prl,aps,epsfig]{revtex}         
\begin{document}
\draft               
\twocolumn[\hsize\textwidth\columnwidth\hsize\csname @twocolumnfalse\endcsname

\title{Spatial solitons in a pumped semiconductor resonator}
\author{V. B. Taranenko, C. O. Weiss, W. Stolz$^*$}
\address{Physikalisch-Technische Bundesanstalt 38116 Braunschweig, Germany\\
$^*$Philipps Universitaet, 35032 Marburg, Germany}
\maketitle
\begin{abstract}
Bright and dark spatial solitons are observed in an optically
pumped semiconductor resonator. The pumping allows to
considerably reduce the light intensity necessary for the
existence of the solitons and alleviates thermal load problems.
Experiments are found to agree with calculations based on a
simple large aperture semiconductor resonator model. The role of
signs of the absorptive and reactive nonlinearities for soliton
existence are discussed in relation with the nonlinear resonance
effect, the tilted wave mechanism of pattern formation, and the
sign of the population inversion.
\end{abstract}
 \vskip1pc ]
Spatial resonator solitons are presently considered for optical
information processing, in the form of mobile binary information
carriers \cite{tag:1}. We have recently shown that such solitons
exist in semiconductor quantum well resonators \cite{tag:2}. The
speed with which these solitons can be switched-on was, however,
found to be limited to $\sim$ 1 $\mu$s, as a consequence of the
high local heat dissipation inside the bright solitons
\cite{tag:2,tag:3}. These thermal effects limit also the speed
with which these solitons can be moved. The high dissipation in
the solitons relates directly to the high light intensity at
which these solitons exist. Typical intensities supporting the
solitons are $\sim$ 1 kW/cm$^{2}$ or the power needed to support
one soliton of $\sim$ 10 $\mu$m size is $\sim$ 1 mW; a sizeable
fraction of which is dissipated.

To reduce the high light intensities needed to support solitons,
and with this to ease the limitations due to dissipation, we
attempt to use a pumped medium rather than a purely absorbing one
inside the resonator. Conceptually the idea being that part of
the energy supporting the solitons be provided incoherently,
reducing the energy to be supplied in a coherent form. For
simplicity and flexibility we have used optical pumping in the
experiment, which could, of course, be replaced by electrical
pumping in the technical applications.\\

The set-up for the observations is largely as described in
\cite{tag:2}. In short, it uses a \textit{cw} Ti:Al$_2$0$_3$-laser
tunable to chose the desired wavelength region - in our case near
the semiconductor band edge. This laser illuminates an area of
$\sim$ 50 $\mu$m diameter on the semiconductor resonator sample,
corresponding to a Fresnel number of $\sim$ 100. The light
reflected from the sample is observed, taking ns-snapshots of the
illuminated area and following the reflected intensity in time in
particular points in the area. Details are given in
\cite{tag:2,tag:3}.

For optical pumping light from a 750 mW diode laser with spectral
width $\sim$ 3 nm, and thus rather incoherent, is spatially
superimposed onto the Ti:Al$_2$0$_3$-laser light. The resonator
sample used, described in \cite{tag:4}, contains 18 quantum wells
and has mirrors of $\sim$ 99.6 $\%$ reflectivity. The optical
pumping wavelength was chosen as 810 nm (about 40 nm shorter than
the band gap wavelength) in the stop band of the resonator mirrors
\cite{tag:4}. The coherent laser radiation near the band edge
wavelength was, as usual for soliton formation, blue-detuned with
respect to the resonator resonance (tilted wave mechanism for
structure formation, see \cite{tag:5}).\\

\begin{figure}[htbf] \epsfxsize=80mm
\centerline{\epsfbox{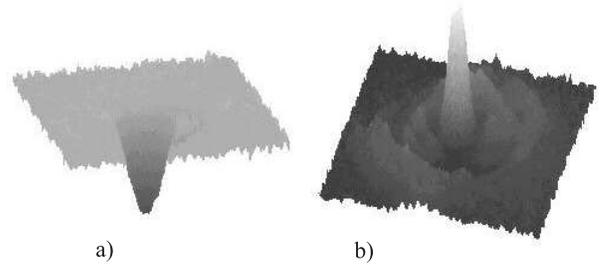}} \vspace{0.5cm} \caption{Spatial
resonator solitons in the pumped semiconductor QW-resonator.
Observation in reflection. a) bright soliton (view from
"bottom"), b) dark soliton (normal or "top"-view). The vertical
coordinate is the sample reflectivity. Pump strength is near the
crossing point of the switching thresholds in Fig.~2.}
\end{figure}

We find that bright and dark spatial resonator solitons exist
throughout the whole range of available pump intensities (see
Fig.~1). Fig.~2 shows the variation of the switch-on and
switch-off intensities of the bistable resonator, with the pump
intensity. It is seen that the light intensities supporting
spatial solitons are reduced at the maximum of the pump power
available here by nearly an order of magnitude, compared to the
unpumped case. This should constitute a significant advantage in
applications. The measurements were routinely carried out
illuminating the resonator sample for a few ms only, to limit
thermal phenomena. The switching thresholds in Fig.~2a were
determined from recordings such as Fig.~4a where the switch-on and
-off intensities are evident. For higher pump intensities Fig.~2a
shows that the switch-on intensities, measured in this way, lie
below the switch-off intensities. We would interpret this as an
artifact due to a residual thermal effect: At the time of
switch-on, the material is "cold". When the material is switched
on, the intensity inside the soliton and with it the dissipation
is high, thus at the time of switch-off the material temperature
is increased, and the band gap, and with it the bistability
characteristic, has shifted. The switch-off intensity is then
increased over the value of the "cold" material and thus the
measured switch-on and switch-off thresholds appear to cross.

\begin{figure}[htbf]
\epsfxsize=88mm \centerline{\epsfbox{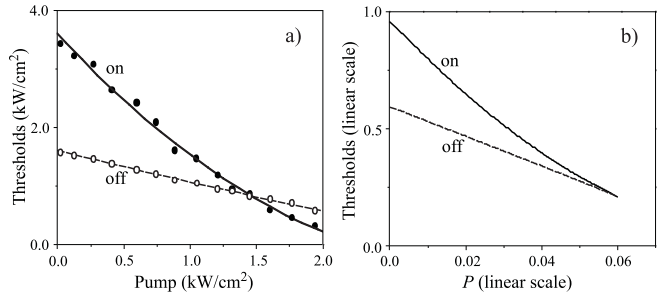}} \vspace{0.5cm}
\caption{a) Measured switch-on and switch-off intensities of the
bistable resonator as a function of pump strength. Solid and
dashed lines given to guide the eye. b) Calculated plane wave
switching intensities obtained from model (1). Parameters are
$\eta=0.5$, $C=36$, $\rm Im(\alpha)=1$, $\rm Re(\alpha)=-0.02$,
$\theta=-2$, $\gamma=0.1$. Transparency corresponds to $P=0.1$.}
\end{figure}

In the presence of pumping the power dissipated locally (in the
soliton) which causes the limitations in switching and movement
speed, mentioned above, is reduced according to the switching
threshold. This can be expected to alleviate the limitations
given by dissipation significantly.

Fig.~2b shows the switching thresholds for plane waves as
calculated from the rate equation model for the intracavity
optical field $E$ and carrier density $N$ (similar to
\cite{tag:6}):

\begin{equation}
\cases{{\partial E}/{\partial t}=E_{\rm in}-E[1+\eta+C{\rm
Im}(\alpha)(1-N)]-\cr
\quad\quad\quad\quad\quad\quad\quad\quad-iE[\theta-C{\rm
Re}(\alpha)N-\nabla^{2}_{\bot}]\,,\cr \cr{\partial N}/{\partial
t}=P-\gamma[N-|E|^2(1-N)-d\nabla^{2}_{\bot}N]\,,}
\end{equation}
where $E$$_{\rm in}$ is the incident field, $\eta$ is the linear
intracavity absorption, $C$ is the bistability parameter
\cite{tag:6}, Im($\alpha$)(1-$N$) and Re($\alpha$)$N$ are used to
describe the absorptive and refractive nonlinearities,
respectively. $\theta$ is the detuning of the optical field from
the resonator resonance, $\gamma$ is the ratio of the photon
lifetime in the resonator to the carrier nonradiative
recombination time. $P$ is the optical pump rate, $d$ is the
diffusion coefficient (normalized to the diffraction coefficient)
and $\nabla^{2}_{\bot}={\partial^{2}\nonumber}/{\partial
x^{2}}+{\partial^{2}\nonumber}/{\partial y^{2}}$ ($x$, $y$ are the
transverse coordinates) accounts for nonlocalities. Figs~2a and
2b show a good qualitative agreement.

Fig.~3 shows actual numeric soliton solutions of (1) as they are
found for pump values close to the point where the on- and
off-thresholds for plane waves (Fig.~2b) intersect (small
hysteresis range). The steady state, plane wave characteristic
(in reflection) of the resonator is given. For the same value of
resonator detuning bright solitons appear closely below the plane
wave switch-on and dark solitons closely above it.

\begin{figure}[htbf]
\epsfxsize=65mm \centerline{\epsfbox{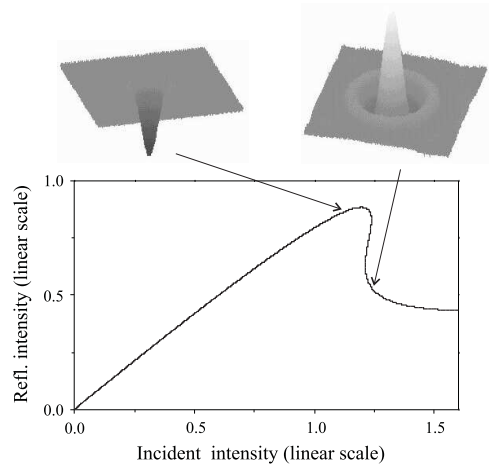}} \vspace{0.5cm}
\caption{Plane wave resonator characteristics calculated from
model (1). Bright and dark soliton solutions shown on top exist
close to the bistability range. Parameters are $\eta=0.5$,
$C=36$, $\rm Im(\alpha)=1$, $\rm Re(\alpha)=-0.02$, $\theta=-2$,
$P=0.055$, $\gamma=0.1$, $d = 0.1$.}
\end{figure}

Fig.~4 shows observation of the dynamic behavior of the bright
and dark soliton as measured under these conditions. Fig.~4a is
the case of bright solitons. The intensity reflected from the
center of the soliton is given. Without pumping the resonator
does not switch and reflects all incident light. For the same
illuminating intensity the soliton switches on with pumping. The
switch-on is apparently direct and fast and not mediated by
thermal effects as for soliton formation without pumping
\cite{tag:2}.

Fig.~4b shows the light reflected for a dark soliton. Without
pumping the sample switches an area given by the illuminating
beam at $\sim$ 0.4 $\mu$s. No soliton forms here. With pumping the
switching appears at a much lower intensity and a dark soliton is
formed because the illuminating intensity is higher than in Fig.
4a. The slow change of intensity measured over the time of
observation reflects a motion of the dark soliton, as has been
shown in \cite{tag:2}. This motion is a consequence of the
temperature difference between the interior and the exterior of
the soliton. Such temperature difference does not allow a dark
soliton to sit stationarily in one location, but forces it to
move continually \cite{tag:7}, much in the way discussed for
optical vortices in class B-lasers \cite{tag:8}. We note here (as
in \cite{tag:2}) that the reflectivity at the soliton center is
larger than 1 (!), indicating that the soliton collects energy
from its surrounding.

In \cite{tag:2,tag:3} we found that under normal conditions the
formation of bright solitons is slowed by a process determined by
the heating due to the high intensity inside the soliton. We note
here that a time-(and space-) independent heating does not
produce such problems. This is evidenced by the direct switch-on
of the soliton in Fig.~4a. The uniform heating due to the pump
causes only a time-independent shift of the bistability
characteristics (together with the basin of attraction for
solitons), altering switching thresholds, but nothing else. Thus
the heating by the pump does not slow soliton formation. Rather
it allows the reduction of the soliton supporting coherent light
intensity, and with it the time- and space-dependent heating.

\begin{figure}[htbf]
\epsfxsize=65mm \centerline{\epsfbox{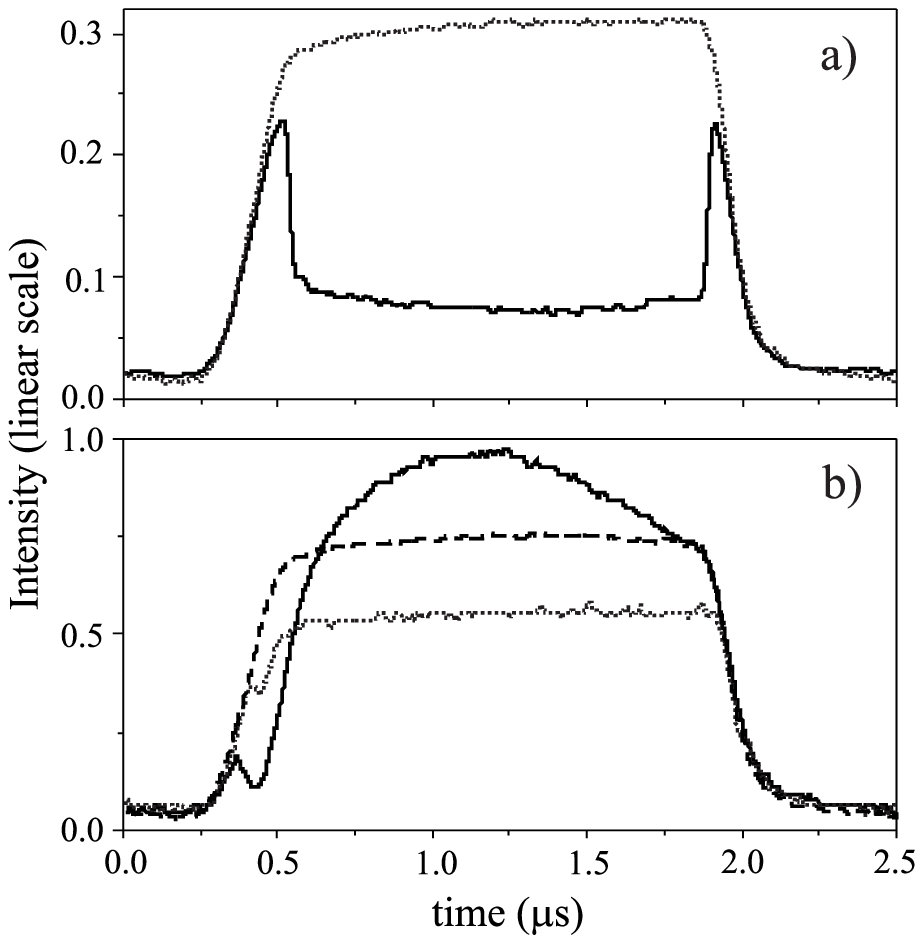}} \vspace{0.5cm}
\caption{Dynamical features of bright soliton (a)) and dark
soliton (b)).\\ a) Dotted line: reflected intensity without
pumping. The switch-on intensity is not reached, thus incident
intensity equals the reflected intensity. Solid line: with
pumping, a soliton switches on directly (without delay from
thermal effects) at 0.5 $\mu$s, and off at a slightly smaller
intensity. Pump strength near the threshold crossing point of
Fig.~2 (as in Fig.~1).\\ b) Dashed line: incident intensity.
Dotted line: reflected intensity without pumping. A certain area
in the illumination beam is switched on at 0.5 $\mu$s, but no
soliton formed. Solid line: reflected intensity with pumping. A
dark soliton is switched on at 0.5 $\mu$s. Slow changes of the
reflected intensity are due to motion of the soliton (see text).\\
Intensity scales of a), b) normalized to the same value. }
\end{figure}

The pumping allows i) to reduce dramatically the light intensity
supporting the solitons, ii) to switch on solitons fast without
thermal delay. This would also imply that with pumping thermal
effects would not limit the speed with which solitons can be
moved.

In these pumping experiments we have worked below population
inversion in the quantum-well-material. This is concluded in the
following way: At the transparency point all material
nonlinearities can be expected to vanish and with them any
bistability. We have not crossed that point with the pump
intensities available to us (Fig.~2) as bistability is found up
to the highest pump value.

For our sample with its high reflectivity mirrors one would also
think that the transparency point and the laser threshold would
be extremely close in pump intensity. The resonator needs only
inversion of 1 $\%$ above transparency to emit as a laser, so that
it would be difficult to keep the pump power in this very narrow
range.

We mention one essential difference in bistability and solitons
below and above transparency. Below transparency the absorptive
nonlinearity (bleaching) has the correct sign for a resonator
bistability, while above transparency (gain saturation) the sign
is wrong. Consequently, below transparency solitons based on the
dissipative part of the nonlinearity can exist, while above such
solitons can not exist. Evidently, below, as well as above
transparency, solitons can be supported by the reactive
(dispersive) part of the nonlinearity.

Additionally concerning the reactive part of the nonlinearity,
there is also a difference below and above transparency. The
nonlinear resonance mechanism of soliton formation \cite{tag:9}
requires (counter intuitively) a defocusing nonlinearity below
transparency and a focusing nonlinearity above transparency for a
blue-detuned resonator. Self-focusing is generally furthering
soliton formation. Working above transparency, would therefore
seem to be advantageous for solitons.\\

In summary, we have shown the beneficial effects of pumping on
the properties of spatial resonator solitons. The necessary light
intensities can be drastically reduced and thermal problems
hindering fast switching and fast motion are reduced. How close
one can work above or below the transparency point, i.e. how
small the coherent intensities supporting solitons can be,
depends evidently on the requirements of soliton stability, as
the existence ranges reduce, as one approaches transparency.
Although we have used optical pumping for convenience here, the
option exists to replace it by electrical pumping in
applications.\\

Acknowledgement\\ We acknowledge beneficial discussions with
D.Michaelis and K.Staliunas. Work supported by ESPRIT under
project number 28235 and Deutsche Forschungsgemeinschaft under
project number we743/12-1.

\end{document}